# Single Defect Center Scanning Near-Field Optical Microscopy on Graphene

J. Tisler, T. Oeckinghaus, R. Stöhr, R. Kolesov, F. Reinhard and J. Wrachtrup

3. Institute of Physics, Stuttgart University, 70550 Stuttgart, Germany

## Abstract

We demonstrate high resolution scanning fluorescence resonance energy transfer microscopy between a single nitrogen-vacancy center as donor and graphene as acceptor. Images with few nanometer resolution of single and multilayer graphene structures were attained. An energy transfer efficiency of 30% at distances of 10nm between a single defect and graphene was measured. Further the energy transfer distance dependence of the nitrogen-vacancy center to graphene was measured to show the predicted $d^{-4}$ dependence. Our studies pave the way towards a diamond defect center based versatile single emitter scanning microscope.

## Introduction

Imaging the optical near field of nano-sized structures is of fundamental interest to various areas of photonic- , material- and biological sciences.

As a result a wealth of scanning near-field optical microscopy (SNOM) techniques are available nowadays [1]. A notable extension to SNOM is scanning fluorescence resonance energy transfer microscopy [2]. The method exploits fluorescence resonance energy transfer (FRET)[3], a non-radiative dipole-dipole interaction between transition dipole moments of a donor and an acceptor. As the method exploits the near field interaction of two induced dipoles it has a steep dependence on distance d ($z^{-6}$) and promises to be capable of achieving nm resolution. Different donors and acceptor systems have been used like dyes[4], color centers[5] or quantum dots[6]. But most of these systems suffer from blinking or limited photostability rendering their use cumbersome in practice At low temperature single molecular fluorescence usually is stable and as a consequence scanning fluorescence imaging of light fields has been demonstrated [7]. One particular photoactive system, namely the nitrogen-vacancy (NV) center in diamond has gained attention as an atom-like photon emitter. It proves to be photostable even at room temperature [8], can be brought into nm proximity of any photonic system [9] and has been proposed as a stable emitter for near-field microscopy[4]. As a result FRET has been demonstrated between a NV and another single molecule with high transfer efficiency [10,11].Therefore it appears to be an ideal emitter for FRET SNOM. In this study we utilize a scanning near field FRET microscope to investigate the interaction of a single NV emitter with graphene. Graphene has a number of intriguing near-field optical properties like high mode density[12] and it is speculated that graphene plasmons are easily lauched by near field interactions with emitters. Further on only 2.3% of the incoming far field plane wave absorbed from a monolayer[13] making near and far field effects to be clearly distinguishable. Owing to its two dimensional geometry the distance dependence of the FRET efficiency is expected to be $z^{-4}$ dependent[14] rather that $z^{-6}$. Additionally the FRET transfer efficiency probes relevant material properties like the local fermi energy.

## Results

All the experiment use a home-build scanning FRET microscope based on an atomic force microscope (AFM) with an optically accessible tip. Single NV centers inside nanodiamonds on the tip of the AFM are used as FRET donors. A laser beam exciting the NV center is focused through the sample onto the NV. Fluorescence of the defect center is collected through the same channel (see Fig. 1).

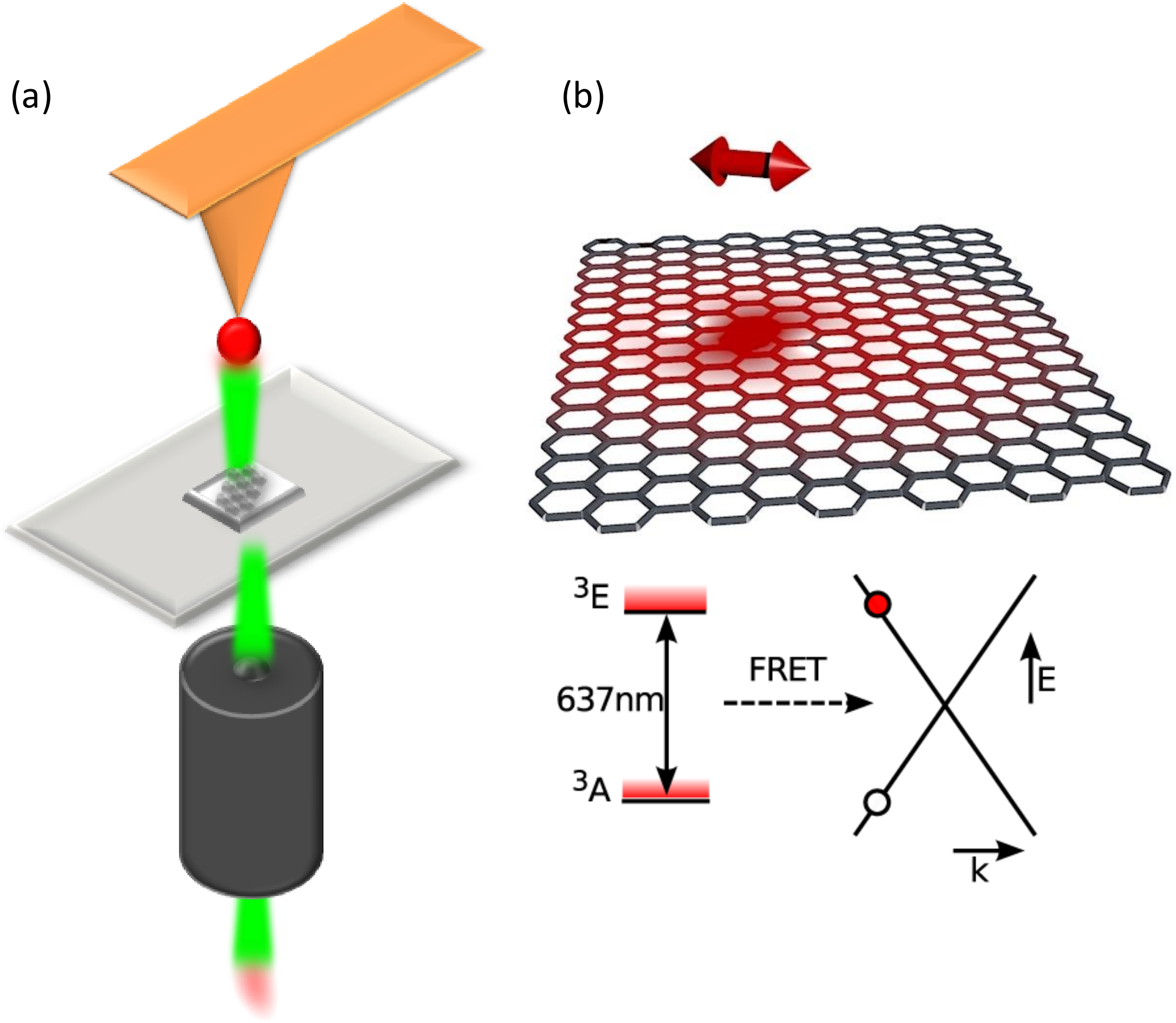

**Figure 1: Schematic images of the experiment (a) Schematic image of the experimental setup. Nanodiamonds are glued onto the apex of a Si tip. A high numerical aperture objective (NA 1.3 is used to excite a single NV in the nanodiamond and collect NV fluorescence. (b) Energy levels of the lowest optical**

**transition relevant for energy transfer and and graphene band structure near the Γ point.**

To allow for close proximity between the defect and graphene small nanodiamond with diameters around ~ 25nm containing single NV centers were used. When the NV center is placed in close proximity to the graphene sample, its fluorescence is quenched by Förster energy transfer. In this process, an electronic excitation of the NV center is nonradiatively transferred into an exciton in graphene (Fig. 1b) which quickly dissipates excitation energy mostly by internal radiationless decay [15] ccuring at a rate $\gamma_{nr}$, this process competes with radiative emission of the NV center and reduces the fluorescence intensity to a value

$$I = \frac{\gamma_{ex}\gamma_r}{\gamma_r + \gamma_{nr}}$$

where $\gamma_{ex}$ and $\gamma_r$ are the excitation rate and radiative rate, respectively, and we assume $\gamma_{ex}$ to be much below saturation.

The quenching rate $\gamma_{nr}$ can be computed from Förster's law[16] with the additional assumption that a sheet of graphene can be approximated by a two-dimensional array of infinitesimal flakes[17].

$$\gamma_{nr} = A \iint_{\substack{\text{graphene}\\ \text{sheet}}} ds^2 \, \left|\boldsymbol{E_p}\right|^2 {\mu_g}^2 = A \iint_{\substack{\text{graphene}\\ \text{sheet}}} ds^2 \, \frac{{\mu_{eg}}^2 {\mu_g}^2 \, f(\hat{\boldsymbol{r}}, \widehat{\boldsymbol{\mu_{eg}}})}{r^6} \quad (1).$$

Here, A is a proportionality constant and $\boldsymbol{E_p} = \boldsymbol{E} - (\boldsymbol{e_z} \cdot \boldsymbol{E})\boldsymbol{e_z}$ denotes the in-plane component of $\boldsymbol{E} = \left(3\hat{\boldsymbol{r}}\left(\boldsymbol{\mu_{eg}} \cdot \hat{\boldsymbol{r}}\right) - \boldsymbol{\mu_{eg}}\right)/(4\pi\epsilon_0 r^3)$, the near-field of the transition dipole $\boldsymbol{\mu_{eg}} = -e\langle e|\boldsymbol{r}|g\rangle$ of the NV center's optical transition between states $|g\rangle$ and $|e\rangle$ (red shade in Fig. 1b). The graphene transition dipole moment $\mu_g$is a scalar, reflecting the fact that graphene is an isotropic material. Hence, exciton transitions can be driven by any in-plane driving field $\boldsymbol{E_p} = E\,\boldsymbol{e_p}$, regardless of its polarization $\boldsymbol{e_p}$ . Precisely, $ds^2\,\mu_g{}^2 = e^2 \sum_{\boldsymbol{k_i},\boldsymbol{k_f},\omega=v_F(\boldsymbol{k_f}+\boldsymbol{k_i})} \left|\langle \boldsymbol{k_f}|x|\boldsymbol{k_i}\rangle\right|^2$ where $|\boldsymbol{k_i}\rangle, |\boldsymbol{k_f}\rangle$ denote plane-wave electron states in graphene and $\omega$ is the NV transition's frequency.

Integrating equation (1) over an infinitely large graphene surface yields a modified Förster type law

$$\gamma_{nr} = \gamma_r \frac{z_0^4}{z^4}, \qquad (2)$$

with a Förster distance $z_0$ and a quenching rate rising with the fourth power of distance, differing markedly from the $z^{-6}$ law commonly known for point like objects such as molecules or atoms.

Quantitatively, the Förster distance can be calculated from eq. (2) and the expressions of $\gamma_{nr}$ and $\gamma_r$ stated in previous work [7,14].

$$z_0 = \sqrt[4]{\frac{9e^2\hbar^3c^3}{4\cdot 512\,(\hbar\omega)^4\epsilon_0}} = 15.3\,\text{nm}. \quad (3)$$

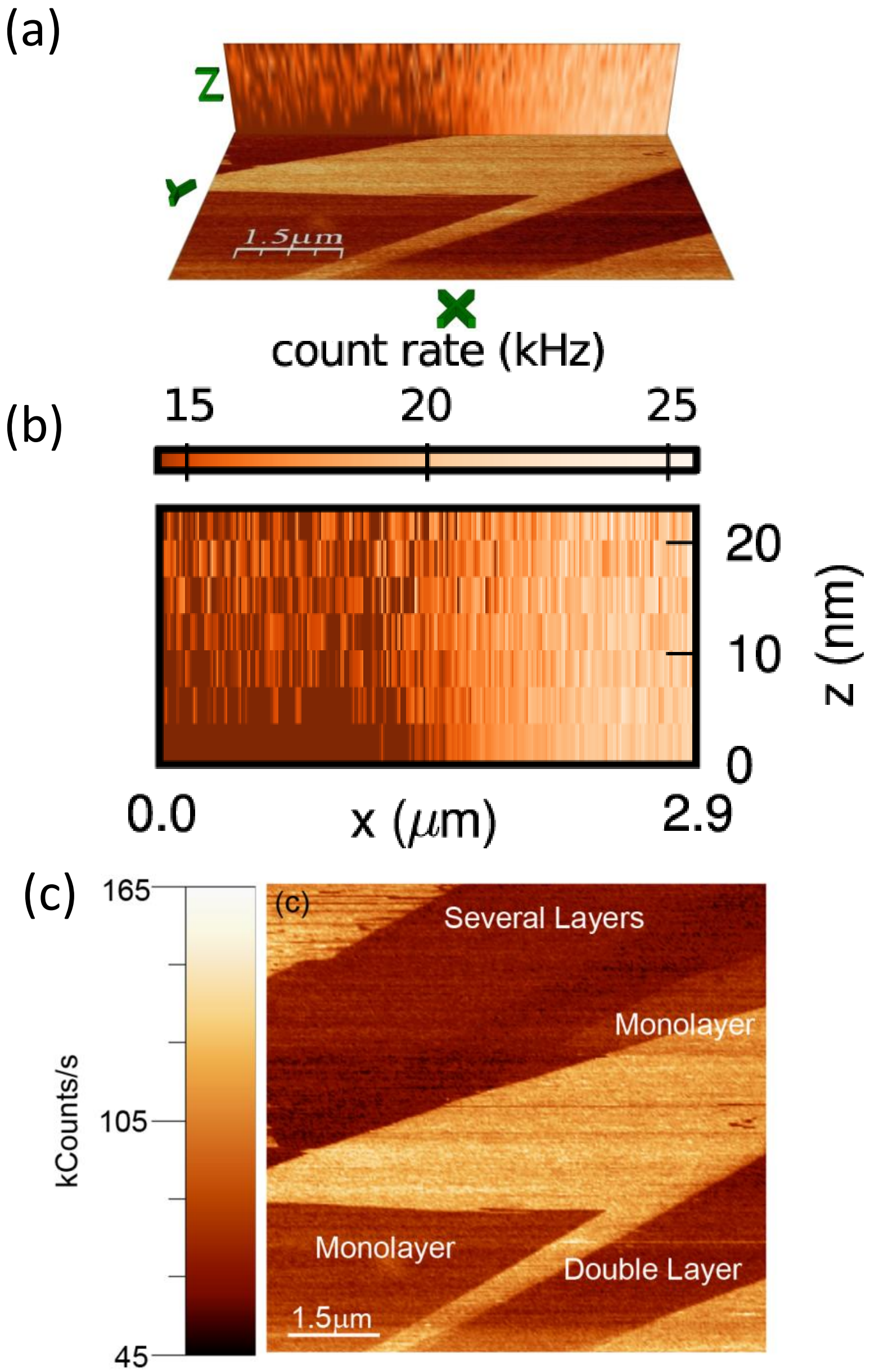

**Figure 2: Scanning FRET on graphene mono- and multiple layers. Plots show the NV fluorescence intensity as a function of lateral (c) as well as lateral and axial position over the graphene sample.**

In the following we discuss our scanning FRET experiments on layers of different graphene thickness. It is worth mentioning that in addition to lateral xy scans we also measured in xz- and in xy-direction. For such images tapping mode scanning FRET is used. In this mode the tip of the AFM is oscillating during scanning in contrast to contact mode. The photons arriving during the scan are measured in registry with the different heights of the oscillating tip.. Fig. 2b shows a corresponding image in the xz plane. To compare eq. (1) with results from Fig. 2c, we transform the fluorescence $I(x, y, z)$ observed in measurements into a quenching rate $\gamma_{nr}(x, y, z)$ by the relation

$$\gamma_{nr} = \left(\frac{I_0}{I(x,y,z)} - 1\right)\gamma_r, \quad (4)$$

where $I_0$ is the unquenched NV fluorescence intensity. In all the following, we assume a radiative rate $\gamma_r = 1/\tau_{NV}$ based on the measured fluorescence lifetime $\tau_{NV} = 13\text{ns}$[18]. Note that this conversion implicitly assumes unity quantum efficiency.

In case of Fig. 2 the monolayer quenching rate was $27.7\text{E}^6$ $\text{s}^{-1}$ for a double layer $33.1\text{E}^6$ $\text{s}^{-1}$ and for several layers $46.2\text{E}^6$ $\text{s}^{-1}$. In contrast to previous work[19] no exact scaling with layer thickness was observed, probably due to background contributions which were not taken into account in our measurements.

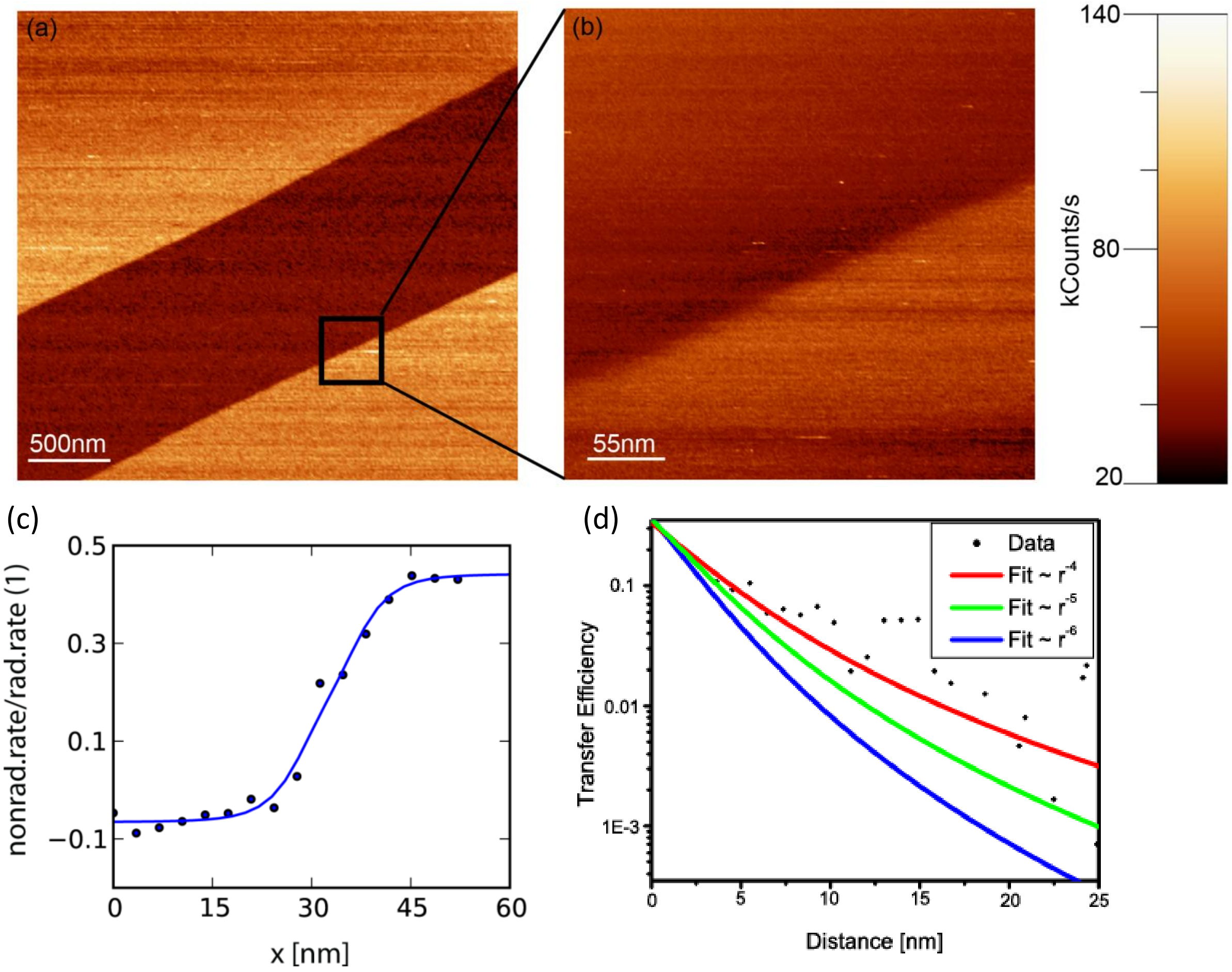

**Figure 3: Quantitative comparison of scanning fluorescence resonance energy transfer microscopy images to theory. A high resolution contact-mode scan of a graphene edge (a-b) is used to measure the step response function of the single NV center scanning over the graphene monolayer (c) The solid line of (c) is a fit to the data using eq. (1), parametrized by the NV-graphene distance of $z_{NV} = 9.8$ nm. (d) Vertical dependence $\gamma_{NV}(z)$. A fit of the data (red line) agrees with the predicted $z^{-4}$ law (eq. (2)) for a Förster distance of $z_0 = 8.25 \pm 0.13$ nm.**

Thanks to their atomic size, scanning single emitters are able to map near-field couplings with molecular (nm) resolution in all three dimensions. As a first application of this remarkable property, we experimentally confirmed the theoretical model of eq. (1). The result is shown in Fig. 3. In the lateral dimensions (Fig. 3 a-c), a high-resolution scan of a graphene edge (Fig. 3b) reveals that fluorescence drops smoothly over a length scale of ~10nm when the NV center is moved over the flake. The theoretical prediction (solid line in Fig. 3c) is obtained by numerically integrating equation (1), taking into account the fact that the NV transition has two orthogonal transition dipoles $\boldsymbol{\mu_1}, \boldsymbol{\mu_2}$

$$\gamma_{nr}(x) = \left(\gamma_{nr}^{\boldsymbol{\mu_1}}(x) + \gamma_{nr}^{\boldsymbol{\mu_2}}(x)\right)/2$$

$$\gamma_{nr}^{\boldsymbol{\mu}}(x) = A{\mu_g}^2 \int_x^{\infty} dx \int_{-\infty}^{\infty} dy \; \left|\boldsymbol{E_p}^{z_{NV},\boldsymbol{\mu}}(x,y)\right|^2.$$

The remaining free parameters $A, {\mu_g}^2$ and $z_{NV}$ are fit to the data and the dipole orientation is chosen as $\boldsymbol{\mu_1} \parallel \boldsymbol{e_z}, \boldsymbol{\mu_2} \parallel \boldsymbol{e_x}$ to best fit the observed bahaviour. We find that the data agrees well with the theoretical prediction (Fig. 3c). The achieved resolution is limited by the spatial extent of the near-field $\boldsymbol{E_p}$, which is of the order of the NV-graphene distance. Reverting the argument, we can extract this distance from the fit, finding that $z_{NV} = 9.8$ nm. This value is nonzero even though the nanodiamond was scanned over the graphene in contact mode. Therefore, we interpret this value as the distance between the NV center and the surface of the nanodiamond.

We also measured the vertical dependence $\gamma_{nr}(z)$ by repeatedly approaching and retracting the tip on a large graphene surface (Fig 3d). Again, we find that the data is well described by eq. (1) and in particular agrees with the predicted $z^{-4}$ law of eq. (2). Using the value $z_{NV}$ obtained from the lateral scan, we can quantitatively fit the data and infer a Förster distance of $z_0 = 8.25 \pm 0.13$ nm . This differs significantly from the theory prediction ($z_0 = 15.3$nm, eq. (3)). Most likely, this difference is due to additional nonradiative channels which lower the NV center's quantum efficiency and thus reduce $z_0$. The existence of such channels is likely and candidates include quenching by the silicon AFM tip or an intrinsically low quantum efficiency in nanodiamonds. Conversely, we note that our result provides a method to measure the quantum efficiency of a single emitter when it is combined with a quantitative theory of excitonic transitions in graphene[1420].

## Conclusion

We demonstrated optical scanning fluorescence resonance energy transfer microscopy with a single NV center in nanodiamond as emitter.  By scanning over a graphene monolayer we attained nanometer resolution images and a transfer efficiency as large as 30%. Furthermore we experimentally confirmed a $z^{-4}$ dependence of the energy transfer rate between a point-like emitter and a graphene monolayer. While graphene is an interesting photonic material in its own right, application of the technique to other nano photonic structures and acceptors, e.g. single molecules certainly would be of great interest. Applications in biological sciences where scanning FRET might become a valuable addition to other FRET based techniques for, e.g. imaging larger protein structures of cellular surfaces are easily envisioned. Such methods may be combined with the magnetic field sensing capabilities of the NV center to yield a truly multifunctional local probe.

**Acknowledgement**

Financial support by the EU via projects SQUTEC and DINAMO as well as the German Science foundation via research group 1493 and SFB/TR 21 is acknowledged.

**References**

1. Betzig, E., Trautman, J. K., Harris, T. D., Weiner, J. S. & Kostelak, R. L. Breaking the Diffraction Barrier: Optical Microscopy on a Nanometric Scale. *Science* **251**, 1468–1470 (1991).
2. Sekatskii, S. K. & Letokhov, V. S. Nanometer-resolution scanning optical microscope with resonance excitation of the fluorescence of the samples from a single-atom excited center. *Jetp Lett.* **63**, 319–323 (1996).
3. Lakowicz, J. R. *Principles of fluorescence spectroscopy*. (Springer: 2006).
4. Gruber, A. *et al.* Scanning Confocal Optical Microscopy and Magnetic Resonance on Single Defect Centers. *Science* **276**, 2012–2014 (1997).
5. Shubeita, G. T. *et al.* Scanning near-field optical microscopy using semiconductor nanocrystals as a local fluorescence and fluorescence resonance energy transfer source. *J Microsc* **210**, 274–278 (2003).
6. Y. Ebenstein, T. M. Quantum-Dot-Functionalized Scanning Probes for Fluorescence-Energy-Transfer-Based Microscopy. (2003).doi:10.1021/jp036135j
7. Michaelis, J. *et al.* A single molecule as a probe of optical intensity distribution. *Opt. Lett.* **24**, 581–583 (1999).
8. Michalet, X. *et al.* Quantum Dots for Live Cells, in Vivo Imaging, and Diagnostics. *Science* **307**, 538 –544 (2005).
9. Kurtsiefer, C., Mayer, S., Zarda, P. & Weinfurter, H. Stable Solid-State Source of Single Photons. *Phys. Rev. Lett.* **85**, 290–293 (2000).
10. Tisler, J. *et al.* Highly Efficient FRET from a Single Nitrogen-Vacancy Center in Nanodiamonds to a Single Organic Molecule. *ACS Nano* **5**, 7893–7898 (2011).
11. Kühn, S., Hettich, C., Schmitt, C., Poizat, J.-P. & Sandoghdar, V. Diamond colour centres as a nanoscopic light source for scanning near-field optical microscopy. *Journal of Microscopy* **202**, 2–6 (2001).
12. Koppens, F. H. L., Chang, D. E. & García de Abajo, F. J. Graphene Plasmonics: A Platform for Strong Light-Matter Interactions. *Nano Letters* (2011).at <http://resolver.caltech.edu/CaltechAUTHORS:20110929-105946388>
13. Nair, R. R. *et al.* Fine Structure Constant Defines Visual Transparency of Graphene. *Science* **320**, 1308–1308 (2008).
14. Swathi, R. S. & Sebastian, K. L. Long range resonance energy transfer from a dye molecule to graphene has (distance)−4 dependence. *The Journal of Chemical Physics* **130**, 086101–086101–3 (2009).

15. Stöhr, R. J., Kolesov, R., Pflaum, J. & Wrachtrup, J. Fluorescence of laser-created electron-hole plasma in graphene. *Phys. Rev. B* **82**, 121408 (2010).
16. Förster, T. Zwischenmolekulare Energiewanderung und Fluoreszenz. *Annalen der Physik* **437**, 55–75 (1948).
17. Stöhr, R. J. *et al.* Super-resolution Fluorescence Quenching Microscopy of Graphene. *ACS Nano* **6**, 9175–9181 (2012).
18. Robledo, L., Bernien, H., Sar, T. van der & Hanson, R. Spin dynamics in the optical cycle of single nitrogen-vacancy centres in diamond. *New Journal of Physics* **13**, 025013 (2011).
19. Chen, Z., Berciaud, S., Nuckolls, C., Heinz, T. F. & Brus, L. E. Energy Transfer from Individual Semiconductor Nanocrystals to Graphene. *ACS Nano* **4**, 2964–2968 (2010).
20. Swathi, R. S. & Sebastian, K. L. Resonance energy transfer from a dye molecule to graphene. *The Journal of Chemical Physics* **129**, 054703 (2008).